\documentclass{elsart}
\usepackage{epsfig}
\usepackage{rotating}

\begin{document}

\begin{frontmatter}
\begin{flushright}
FREIBURG-EHEP-97-02
\end{flushright}
\date{25 June 1996}
\title{Analysis of trapping and detrapping in semi-insulating GaAs detectors}
\author{M.Rogalla, Th.Eich, N.Evans, R.Geppert, R.G\"oppert,}
\author{R.Irsigler, J.Ludwig, K.Runge, Th.Schmid, D.G.Marder}
\address{Albert-Ludwigs-Universit\"at Freiburg, Fakult\"at f\"ur Physik, D-79104 Freiburg im Breisgau}

\begin{abstract}
To investigate the trapping and detrapping in SI-GaAs particle detectors we
analyzed the signals caused by 5.48 MeV alpha particles with a charge sensitive preamplifier.
From the bias and temperature dependence of
these signals we determine the activation energies of two electron traps.
Additional simulation and measurements of the lifetime as a function of
resistivity have shown that the EL2$^{+}$ is the dominant electron trap in
semi-insulating GaAs.
\end{abstract}
\end{frontmatter}

\section{Introduction}

GaAs Schottky diodes made of commercially available undoped semi-insulating
(SI) LEC (Liquid Encapsulated Czochralski) material work well as radiation
detectors. However, many studies on SI-GaAs detectors denote an incomplete
charge collection efficiency (CCE)\cite{review}. This signal loss seems to
stem from carrier trapping due to deep levels defects. In addition the CCE
is further reduced due to an incomplete penetration of the electrical field
at low bias voltages\cite{nimalt}. In the first part of this paper we
analyze the shape of signals caused by 5.48 MeV alpha particles ($^{241}$Am)
as a function of bias voltage and temperature for low, medium and high ohmic
SI-GaAs substrates. In the second part an attempt has been made to explain
the variation of the charge collection efficiency on the resistivity of the
materials, which was also reported from B. Berwick et al.\cite{berwick}.

\section{Experimental Method}

The detectors studied in the present work were made on undoped SI-GaAs 3''
wafer obtained from various manufactures. The detectors are 2 or 3 mm in
diameter, 200 $\mu $m thick with circular Schottky pads. The back contact
(Schottky or Ohmic) covers the entire wafer. The details of the contacts are
described elsewhere\cite{roland}. The resistivity was determined from the
I-V characteristic using the Norde plot\cite{norde}.

Charge collection efficiency was measured for irradiation with alphas ($%
^{241}$Am). The spectroscopy chain includes an Vitrom (559-064)%
\footnote{Deutsche Vitrom GmbH, Pinneberg, Germany}
charge sensitive preamplifier with a rise time of 10 ns and a decay time of
240 $\mu $s followed by an ORTEC 579 amplifier-shaper with gaussian shaping
and a time constant of 500 ns. For trapping and detrapping measurements the
output of the preamplifier was read out with a digital scope with 500M
samples/s and band width of 100 MHz. For temperature dependent measurements
the Schottky diodes are placed in an oven with a temperature stability of
about 0.1$^{\circ }$K.

\section{Results and Discussion}

Figure 1 shows the shape of the measured signals from of GaAs detectors
irradiated with alphas from the Schottky contact, applying a positive
voltage to the back contact. The signal is than (Ramo's theorem \cite
{cavallini}) dominated by the motion of electrons rather than holes. We
observe signals similar as reported by ref. \cite{nava} with a fast rise
time and a long exponential decay. The height of the fast part of this
signal increases with increasing bias voltage. The dependence of the time
constant of the slow exponential decay $\tau $ as a function of bias can be
seen in figure 2. If we interpret the slow components as a detrapping from
defects the decrease of $\tau $ can be explained by field enhanced emission
(Poole-Frenkel effect \cite{schroeder}). The observed dependence of $\tau $
on the resistivity $\rho $ is in contradiction to the circuit model proposed
by ref.\cite{nava} where $\tau $ is given by 
\begin{equation}
\tau =\rho \varepsilon \frac Lw
\end{equation}
where $L$ and $w$ denotes the thickness of detector and space charge region.
Additional we have shown in a previous paper \cite{t1} that the space charge
density beyond the Schottky contact increases with increasing resistivity
which results in a small space charge region and increasing electrical field
at the same bias voltage. Taking into account this variation of the electric
field the decrease of the detrapping time as a function of resistivity can
also be explained by the Poole Frenkel effect. If we have a closer look on
the amplitude of the slow component we observe a decrease with bias and
resistivity (figure 3). From temperature dependent measurements at a
constant bias voltage assuming a similar theory as for PICTS (Photo Induced
Current Spectroscopy) \cite{schroeder} the detrapping time is given by

\begin{equation}
\tau =\frac{\exp (E_c-E_T/kT)}{\gamma _n\sigma _nT^2}
\end{equation}
with an activation energy $E_T$, capture cross section $\sigma _n$ and
coefficient $\gamma _n$, we have determined two electron traps. The
activation energies are E$_{T1}$ = 0.352 $\pm $ 0.025 eV and E$_{T2}$ = 0.51 
$\pm $ 0.07 eV (figure 4). A determination of the capture cross section is
not possible because of the time shift caused by Poole Frenkel effect. From
comparison with the values given in the literature the E$_{T1}$ and E$_{T2}$
can be associated with the EL6 and EL3 (table 1)\cite{dcs}.

\begin{table}[tbph]
\caption{Measured activation energies in comperison with the values given in
the literature.}{\small 
\begin{tabular}{|l|l|l|l|l|}
\hline
Trap & Activation energy & Activation energy & Capture cross section & 
Concentration \\ 
& (eV) this paper & (eV) lit. & (cm$^2$) lit. & 10$^{15}$ (cm$^{-3}$) lit.
\\ \hline
EL6 & 0.352 $\pm $ 0.025 & 0.32-0.33 & 2.0$^{.}$10$^{-14}$ & 1-20 \\ \hline
EL3 & 0.051 $\pm $ 0.07 & 0.58 & 0.8-1.7$^{.}$10$^{-13}$ & 0.5-2.0 \\ \hline
\end{tabular}
}
\end{table}

The capture cross sections of both defects are in the 10$^{-13}$ - 10$^{-14}$
cm$^{-2}$ range and the concentrations in melt grown GaAs are in the range
of 0.5$^{.}$10$^{14}$ to 2$^{.}$10$^{16}$ cm$^{-3}$ as reported by M.
Skowronski \cite{dcs}. After the detrapping of the EL6 and EL3 we still not
have 100\% CCE, this means there is still a signal loss due to trapping.
This trap must have a capture cross section and concentrations in the same
range as the two defects we observed directly and a slower detrapping time
constant. If we assume a similar capture cross section and the same
influence of the Poole Frenkel effect on the emission, the activation energy
must be larger then for the measured EL3, because if the energy is smaller
we will expect a detrapping in the temperature range we measured, with a
time constant of a few $\mu $s. The EL2 has an activation energy of 0.8 eV,
the concentration of the ionized state is in the range of 10$^{15}$ cm$%
^{^{-}3}$ \cite{t2} and for electric fields higher than 10$^4$ V/cm a
capture cross section of 8$^{.}$10$^{-14}$ cm$^2$ is published by G. Martin
et al. \cite{dcs1}. This good agreement with claimed properties of the deep
level and also the observed influence of the CCE on the EL2$^{+}$
concentration \cite{grenoble} favour the EL2 to be the dominating electron
trap in SI-GaAs. Therefore we analyze in the following the CCE as a function
of the EL2$^{+}$ or rather the resistivity.

Figure 5 shows the CCE of detectors obtained with low, medium and high ohmic
SI-GaAs substrates. The CCE as a function of bias voltage rises rapidly and
saturates for high voltages. Simulations \cite{nimalt} and measurements \cite
{5} of the field penetration as a function of bias voltage have been shown a
linearly increase of the width of the high field region with bias. From this
we can conclude that the observed saturation of the CCE is due to the
limitations caused by the mean free drift length of electrons. For electric
fields higher than 10$^4$ V/cm \cite{nimalt} we can assume a constant drift
velocity and there for a constant mean free drift length. Under this
assumption we calculate the mean free drift length of the electrons $\lambda
_e$ using

\begin{equation}
CCE=\frac{\lambda _e}d\left[ 1-\exp \left( -\frac{d-x_0}{\lambda _e}\right)
\right] +\frac{\lambda _h}d\left[ 1-\exp \left( -\frac{x_0}{\lambda _h}%
\right) \right]
\end{equation}

where $\lambda _e$ and $\lambda _h$ denotes the drift lengths of electrons
and holes, $d$ the detector thickness and $x_0$ represents the generation
point of electron hole pairs. Estimated a drift velocity $v_{drift}$ of 1$%
^{.}$10$^7$ cm/s we can extract the lifetime $\tau _{e}$of the
electrons from the plateau value of the CCE at high voltages using 
\begin{equation}
\tau _e=\frac{\lambda _e}{v_{drift}}.
\end{equation}
As shown above there is no significant signal loss to the EL3 and EL6 at
this condition which means the lifetime is independent of this two traps.

The electron lifetime as a function of resistivity is shown in figure 6. We
observed a fast decrease from about 45 ns for a resitivity of 0.4$^{.}$10$^7$
$\Omega $cm down to 0.8 ns at 8.9$^{.}$10$^7$ $\Omega $cm. To investigate
the influence of the EL2$^{+}$ on the electron lifetime we calculate in the
first step from the resistivity the position of the Fermi level in the bulk
material assuming a mobility of 8000 $\frac{cm^2}{Vs}$ for electrons and 380 
$\frac{cm^2}{Vs}$ for holes using: 
\begin{equation}
n=\frac 1{q\rho \mu _e}+\sqrt{\left( \frac 1{q\rho \mu _e}\right) ^2-\frac{%
n_i^2}{\mu _e\mu _h}}
\end{equation}
and 
\begin{equation}
E_F=\frac{kT}q\ln \left( \frac n{N_v}\right)
\end{equation}
Knowing the Fermi level position E$_F$, the energy niveau of the level E$%
_{EL2}$ = 0.69 eV \cite{EL2}, the concentration N$_{EL2}$ = 1.2-1.8$^{.}$10$%
^{16}$ cm$^{-3}$ (typical for LEC material \cite{grenoble}) and electronic
degeneracy g = 0.84 \cite{EL2} we determine the ionized density of the EL2
trap according to 
\begin{equation}
N_{EL2^{+}}=N_{EL2}\left( 1-\frac 1{1+g^{-1}\exp \left( \frac{q(E_{EL2}-E_F)%
}{kT}\right) }\right)
\end{equation}
The change of the ionization due to the formation of the space charge region
can be neglected, because the space charge density is in the range of 10$%
^{11}$-10$^{12}$ cm$^{-3}$ in comparison to N$_{EL2^{+}}$ $\sim $ 10$^{14}$%
-10$^{15}$ cm$^{-3}$.

Under the assumption that one single trapping center is prevailing, the
electron lifetime should be inversely proportional to the density of ionized
EL2 (Schockley-Read-Hall statistics): 
\begin{equation}
\tau =\frac 1{\sigma \left\langle v_{th}\right\rangle N_{EL2^{+}}}=\frac
1{\sigma \left\langle \sqrt{\frac{3kT}{m_{eff}}}\right\rangle N_{EL2^{+}}}
\end{equation}
\[
\]
with $\sigma _n$ being the capture cross section and $\left\langle
v_{th}\right\rangle $ the mean thermal velocity. Because of the high
electric field we must take into account the increase of the effective
electron mass $m_{eff}$ due to field enhanced occupation of the second
minimum in the conduction band of GaAs, which results in a lower thermal
velocity than without electric field. For the capture cross section we use
the above mentioned value of 8$^{.}$10$^{-14}$ cm$^2$. Table 2 summarizes
all parameters of the simulation.

\begin{table}[tbph]
\caption{Parameters of the simulation.}
\begin{tabular}{|l|l|l|}
\hline
Parameter & Symbol & Value \\ \hline
Concentration of the EL2 & N$_{EL2}$ & 1.6$^{.}$10$^{16}$ cm$^{-3}$ \\ \hline
Energy level of the EL2 & E$_{EL2}$ & 0.69 eV \\ \hline
degeneracy & g & 0.84 \\ \hline
Temperature & T & 296 K \\ \hline
capture cross section & $\sigma $ & 8$^{.}$10$^{-14}$ cm$^2$ \\ \hline
effective electron mass & m$_{eff}$ & 1.2 \\ \hline
electron mobility & $\mu _e$ & 8000$\frac{cm^2}{Vs}$ \\ \hline
hole mobility & $\mu _h^{}$ & 380$\frac{cm^2}{Vs}$ \\ \hline
\end{tabular}
\end{table}

Using this parameters we obtain a good agreement with the experimental data
(figure 6). This simulation also explains the decrease of the slow signal
height of the detrapping component at low voltages as a function of
resistivity. For material with a short EL2$^{+}$ trapping time more
electrons are captured by the EL2 and less can be trapped by the EL6 and
EL3. In the other case of low resitivity material with only a small
concentration of EL2$^{+}$ the signal loss due to the EL6 and EL3 exceed
that of the EL2$^{+}$.

\section{Conclusion}

We report an new technique to investigate detrapping in SI-GaAs particles
detectors, which is similar to PICTS measurements. This technique allows to
determine directly the electron traps responsible to a signal loss in
particle detectors. Therefore we have identified the EL6 and EL3 as electron
trapping centers. A comparison of measurements and results of a simple
simulation shows that the EL2$^{+}$ is the dominate electron trapping center
in SI-GaAs and responsible for the dependence of the electron lifetime on
the resitivity of the materials.

\section{Acknowledgments}

This work has been supported by the BMBF under contract 057FR11I.

\newpage\ 

\begin{figure}[htbp]
   \begin{center}
\begin{turn}{270}
      \mbox{
          \epsfxsize=10 cm
           \epsffile{fig1.eps}
           }
           \end{turn}
   \end{center}
\caption{
 Measured output signals from GaAs detectors irradiated with alpha
particles on the front for different bias voltages.
}
\end{figure}

\begin{figure}[htbp]
   \begin{center}
\begin{turn}{270}
      \mbox{
          \epsfxsize=10cm
           \epsffile{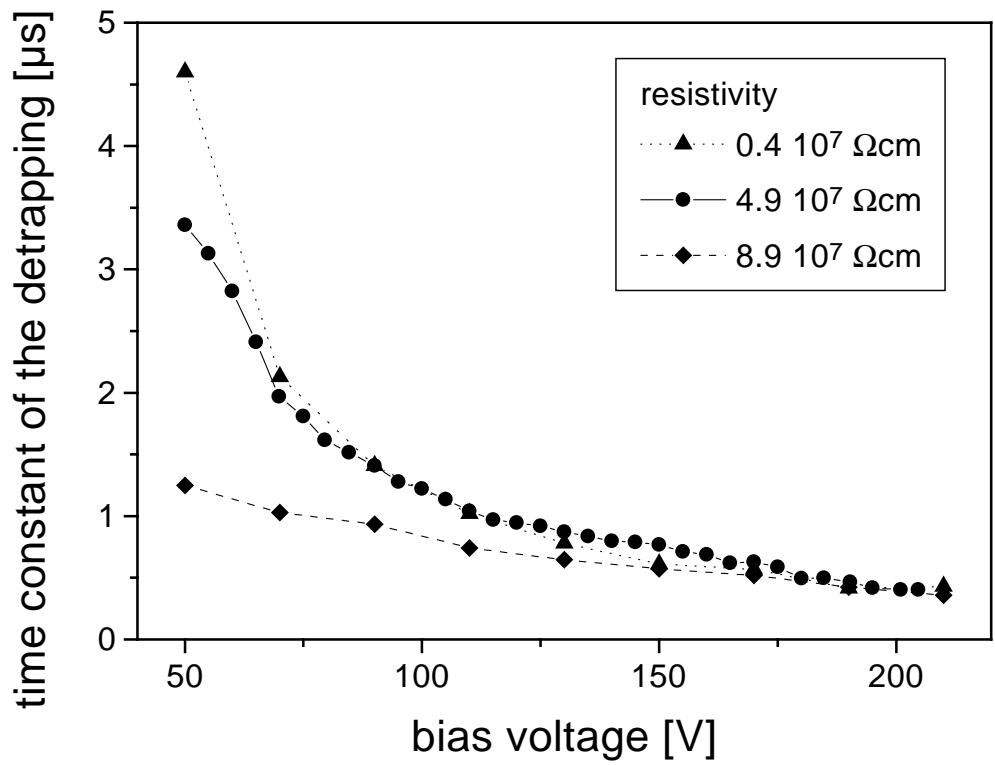}
           }
            \end{turn}
   \end{center}
\caption{
 The time constant of the slow exponential decay as a function of
the bias voltage for different materials.
}
\end{figure}

\begin{figure}[htbp]
   \begin{center}
\begin{turn}{270}
      \mbox{
          \epsfxsize=10cm
           \epsffile{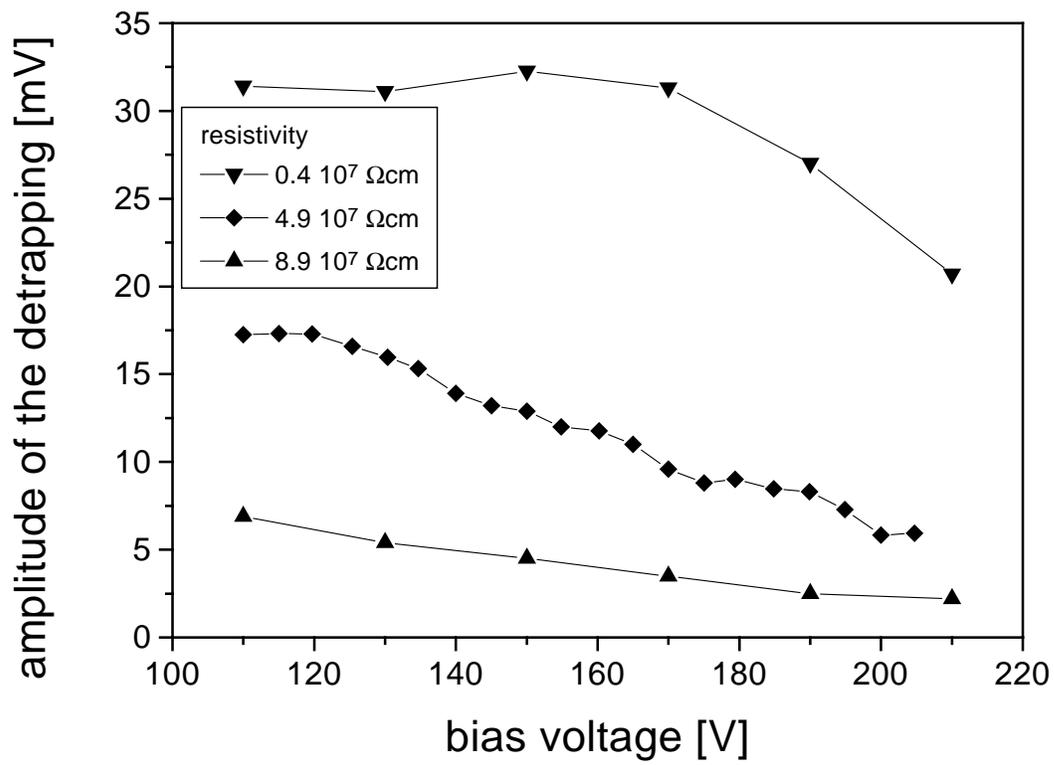}
           }
            \end{turn}
   \end{center}
\caption{
Amplitude of the slow exponential decay as a function of bias
voltage for different materials.
}
\end{figure} 

\begin{figure}[htbp]
   \begin{center}
\begin{turn}{270}
      \mbox{
          \epsfxsize=10cm
           \epsffile{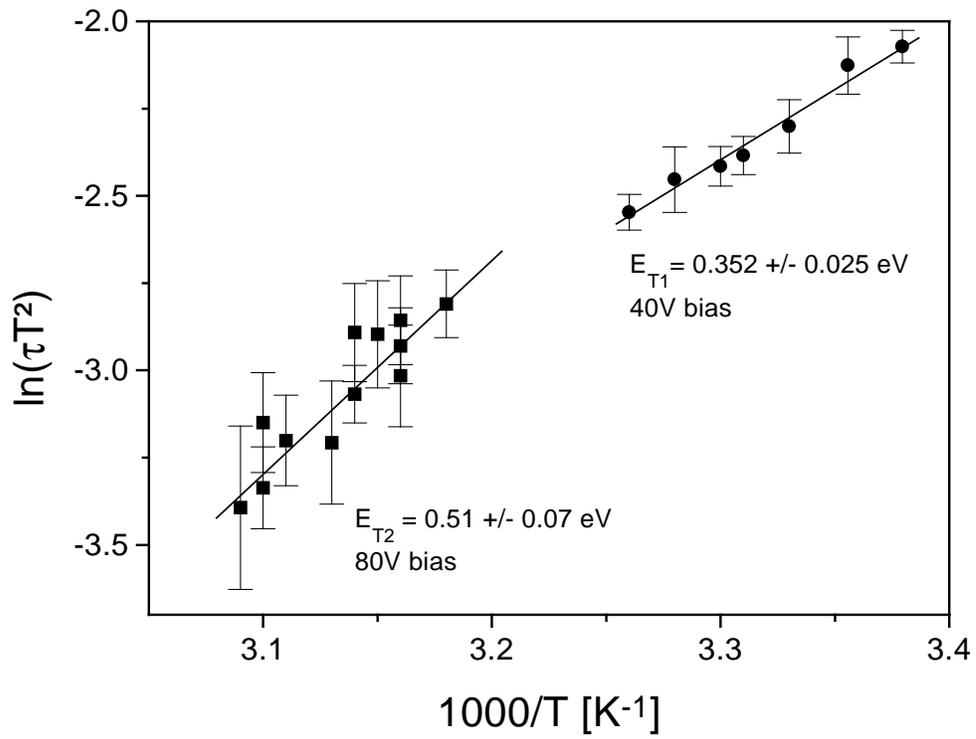}
           }
            \end{turn}
   \end{center}
\caption{
Arrhenius plots of the time constant of the slow exponential
decays measured at bias voltages of 40 V and 80 V (MCP90).
}
\end{figure}

\begin{figure}[htbp]
   \begin{center}
\begin{turn}{270}
      \mbox{
          \epsfxsize=10cm
           \epsffile{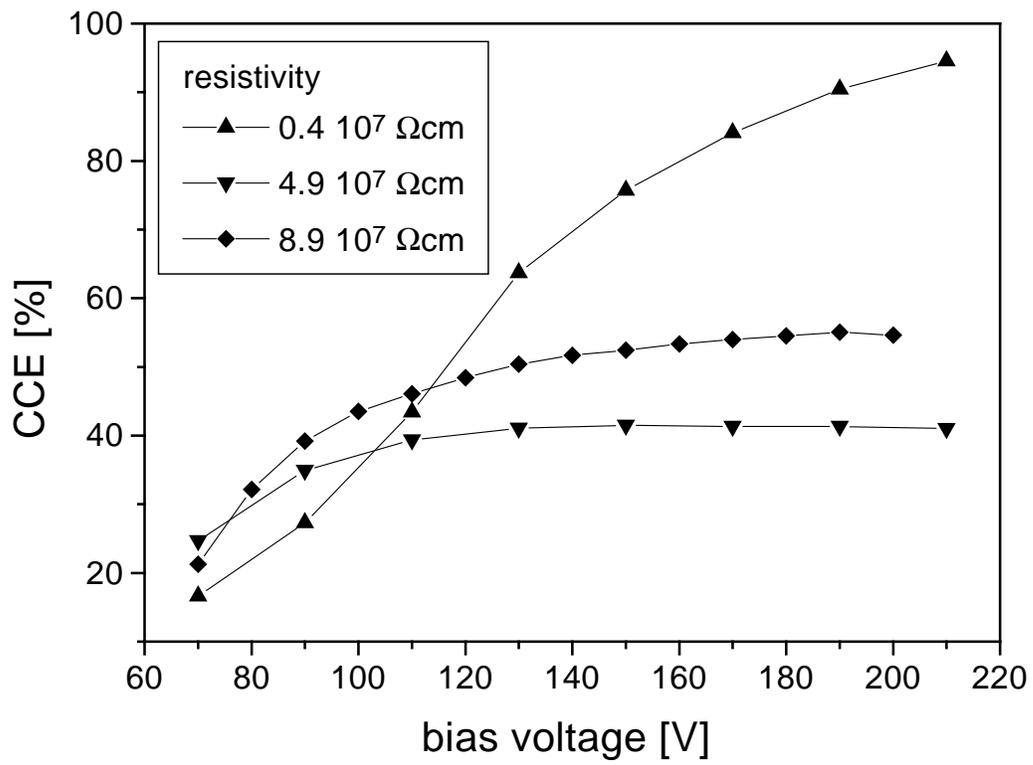}
           }
            \end{turn}
   \end{center}
\caption{
 The charge collection efficiency as a function of the bias voltage
for different materials.
}
\end{figure}

\begin{figure}[htbp]
   \begin{center}
\begin{turn}{270}
      \mbox{
          \epsfxsize=10cm
           \epsffile{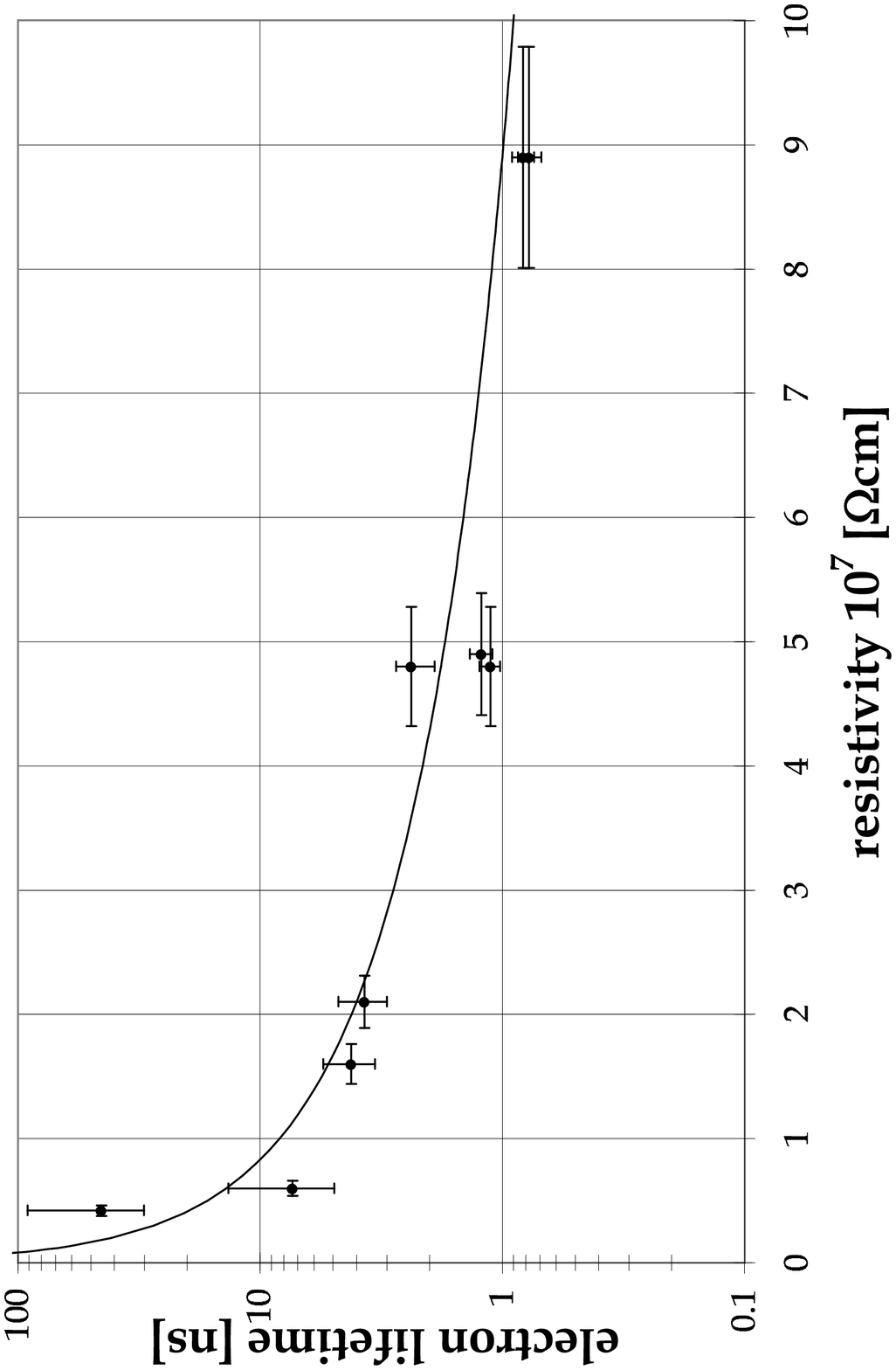}
           }
            \end{turn}
   \end{center}
\caption{
 Comparisons of the measured and simulated dependence of the
electron lifetime on the bulk resistivity.
}
\end{figure}

\end{document}